\documentclass[11pt]{llncs}
\usepackage{calc}
\usepackage{tikz}
\usetikzlibrary{automata,decorations,shapes}
\usetikzlibrary{chains}
\usepackage{shuffle}  \author{John Maraist}
\institute{Computer Science Department, University of Wisconsin-La
  Crosse, \\ 1725 State Street, La Crosse, Wisconsin 54601, USA,
  \email{jmaraist@uwlax.edu}}
\newtheorem{algorithm}{Algorithm}
\date{\today} \title{Generalized LR parsing and the shuffle operator}
\begin{document}
\maketitle
\begin{abstract}
  We adapt Tomita's Generalized LR algorithm to languages generated by
  context-free grammars enriched with a shuffle operator.  The change
  involves extensions to the underlying handle-finding finite
  automaton, construction of parser tables, and the necessary
  optimizations in constructing a deterministic parser.  Our system is
  motivated by an application from artificial intelligence plan
  recognition.  We argue for the correctness of the system, and
  discuss future extensions of this work.
\end{abstract}
\section{Introduction}
\label{sec:orgheadline6}
In this paper we study the extension of context-free grammars (CFGs)
with \emph{shuffle operations}.  Shuffle operations combine strings so
that the order of symbols from each string is preserved, but
interleaving of the shuffled strings is possible.  For example, both
\literal{m12np3r45} and \literal{mn123pr45} are shufflings of
\literal{mnpr} and \literal{12345}, but \literal{mp12nr345} is not
because the \literal n and \literal p occur in a different order.
Although shuffling has long been an aspect of concurrent systems
analysis, the study of shuffled or \emph{intermixed} languages has
lagged behind their use \cite{restivo:lata2015}.  Work through now on
intermixed languages has focused on the theoretical general properties
of language classes, or on practical approaches to regular expressions
and languages with
shuffling~\cite{sulzmann-thiemann:lata2015,sulzmann-thiemann:jcss-lata2015}.

Our interest in the shuffling operator is motivated by an
application from artificial intelligence.  \emph{Plan recognition}
is the problem of determining the goal (or goals) and plan of an
actor from a sequence of observed actions.  The connection between
plan recognition and parsing is well known \cite{vilain:aaai90}.
The requirements and structure of plan recognition and parsing
algorithms differ in some significant ways.  Grammars enumerate an
order among all structures in a language, and parsers expect that
total order; the libraries defining the plans corresponding to a
goal will often give only a partial order, or no order, in some
cases.  Parsers are directed to the understanding of a stream of
inputs corresponding to a single top-level entity; plan recognizers
should be able to recognize the pursuit of multiple goals at the
same time.  And parsers are designed with the assumption that the
entire string to be parsed is available from the start of parsing,
but plan recognizers are typically expected to draw preliminary
conclusions as soon as each piece of input is available.

Early approaches to plan recognition stayed close to parsing
algorithms; they did not address recognition of multiple goals
executed simultaneously, and they did not accept plan libraries
specifying a non-total order among their steps
\cite{pynadath-wellman:uai-2000,vilain:aaai91}.  Goldman, Geib and
Miller's system PHATT relaxed these restrictions
\cite{geib-goldman:ai-2009,goldman-geib-miller:UAI99}, and subsequent
work significantly improved PHATT's
performance \cite{geib-maraist-goldman:icaps08,mirsky:ijcai2016}.
Geib has also similarly adapted parsers for combinatory categorial
grammars (CCGs), a more complicated representation than CFGs, into
plan recognizers \cite{geib:ijcai-2009}.  These approaches are all
somewhat \emph{ad hoc}, in that they develop a algorithm inspired by a
CFG parsing algorithm but neither address the shuffle operator
explicitly, nor identify its impact on the underlying parser.
Moreover, these approaches are all fundamentally based on
\emph{top-down} parsing algorithms, producing a plan recognizer which
must separately represent the different interpretations of inputs.

Our main contribution here is to extend Tomita's Generalized LR (GLR)
parser \cite{tomita-ijcai-1985,tomita-genlr-book-1991} to languages
generated by CFGs enriched with the shuffle operator.  First, in
Section~\ref{sec:cfsg} we formalize our notion of context-free shuffle
grammars (\grammarclass), and define rewriting relations for
\grammarclass s.  We then introduce our \parseralg\ algorithm in two
steps.  In Section~\ref{sec:nd-glr-s} we present a nondeterministic
\parseralg\ parser, and in Section~\ref{sec:impl} we discuss how the
parser can be efficiently implemented.  Finally we conclude with a
discussion of future research directions.

\section{Context-free shuffle grammars}
\label{sec:cfsg}

We formalize our extension of CFGs as follows: a \grammarclass\ is a
quintuple $(V, \Sigma, R,$ $P, S)$ where \(V\) and \(\Sigma\) are
finite, disjoint \emph{alphabets} of respectively nonterminal and
temrinal symbols; \(R\) is a finite relation associating nonterminals
with strings of nonterminals and terminals, \(R\subseteq
V\times(V\cup\Sigma)^*\); \(P\) is a finite relation associating
nonterminals with sets of nonterminals and terminals, \(P\subseteq
V\times\mathcal P(V\cup\Sigma)\); and \(S\in V\) is the \emph{starting
  symbol}. \(R\) and \(P\) are the rules by which nonterminal symbols
may be rewritten to produce (over possibly several rewrites) terminal
strings.  \(R\) gives the tradtional production rules of CFGs; the
definition of a standard CFG is just a quadruple \((V,\Sigma,R,S)\).
\(P\) gives rules for applying the shuffle operator.  For simplicity
we require that if $(a_0\expTo a_1\shuffle\cdots\shuffle a_n)\in P$,
then each $a_i\not\in\mathrm{dom}(P)$.  For example we might have
rules
\[
   \literal S \expTo\literal{T\shuffle U}
\hspace*{13mm} \literal T \expTo\literal{Wpr}
\hspace*{13mm} \literal U \expTo\literal{12345}
\hspace*{13mm} \literal W \expTo\literal{mn}
\]
corresponding to our earlier example, and so formally the sets
$R=\literal{\{S\expTo Wpr,}$ $\literal{T\expTo 12345, W\expTo mn\}}$ and
$P=\literal{\{S\expTo\{T,U\}\}}$.
We use upright sans-serif script when writing literal examples, as
above, and italicized letters to represent variables or unknown
values, with $a$ ranging over single symbols, $s$ ranging over single
nonterminals, and $u$, $v$ over strings of symbols.
\begin{figure}[t]
  \textbf{Rewriting $\rewrites$}
  \begin{list}{}{
      \settowidth{\labelwidth}{\rrule5}
      \setlength{\leftmargin}{\labelwidth+2\labelsep}}
    \item[\rrule1] If \(a\expTo u\in R\), then \(a\rewrites u\).
    \item[\rrule2] If \(a\expTo\{a_{1},\ldots,a_{n}\}\in P\) then we
      have \(a\rewrites \parterm a{a_1,\ldots,a_n}\).
    \item[\rrule3] If \((\forall 0\leq i\leq n) v_i\in\Sigma^*\) and
      \(v\) is a shuffling of the \(v_i\), \(v\in
      v_0\shuffle\cdots\shuffle v_n\), then \(\parterm
      a{v_0,\ldots,v_n} \rewrites v\).
    \item[\rrule4] If \(u\rewrites u'\), then for any strings
      \(u_0,u_1\), \(u_0uu_1\rewrites u_0u'u_1\).
    \item[\rrule5] If \(u\rewrites u'\), then for any
      strings \(u_0,\ldots,u_m\) and nonterminal $a$,
      \(\parterm a{u_0,\ldots,u_n,u}\rewrites\parterm
      a{u_0,\ldots,u_n,u'}\).
  \end{list}
  \textbf{Marked rightmost rewriting $\Mrewrites$}
  \begin{list}{}{
      \settowidth{\labelwidth}{\mrrule5}
      \setlength{\leftmargin}{\labelwidth+2\labelsep}}
  \item[\mrrule1] If \(a\expTo u\in R\), then \(a\posmark\Mrewrites
    u\posmark\).
  \item[\mrrule2] If \(a\expTo\{a_{1},\ldots,a_{n}\}\in P\) then we
    have \(a\posmark\Mrewrites \parterm a{a_1\posmark,\ldots,a_n\posmark}\).
  \item[\mrrule3] If \((\forall 0\leq i\leq n) v_i\in\Sigma^*\) and
    \(v\) is a shuffling of the \(v_i\), \(v\in
    v_0\shuffle\cdots\shuffle v_n\), then \(\parterm a{\posmark
      v_0,\ldots,\posmark v_n} \Mrewrites\posmark v\).
  \item[\mrrule4] If \(u\Mrewrites u'\), then for any
    \(u_0\in(V\cup\Sigma)^*\) and \(u_1\in\Sigma^*\),
    \(u_0uu_1\Mrewrites u_0u'u_1\).
  \item[\mrrule5] If \(u_i\Mrewrites u_i'\) then
    \(\parterm{a_0}{u_0,\cdots,u_i,\cdots,u_n}\Mrewrites\parterm{a_0}{u_0,\cdots,u'_i,\cdots,u_n}\).
  \item[\mrrule6] For any $s\in\Sigma$, $s\posmark\Mrewrites\posmark s$.
  \end{list}
  \caption{The rewriting and rightmost rewriting relations. }
  \label{fig:rewrite}
\end{figure}

Figure~\ref{fig:rewrite} shows the rules for rewriting in
\grammarclass{s}.  The rewriting relation $\rewrites$ applies to a
pair of strings where the latter is derived from the former by
expanding one nonterminal.  For rewriting according to the expansions
in $R$ we have \rrule1.  Rules \rrule2 and \rrule3 treat shuffle
expressions.  The commas and curly braces introduced in the right-hand
side expansion of \rrule2\ are all interim symbols, not part of \(V\)
or \(\Sigma\), which we use as delimiters in non-final rewritten
strings.  They disappear when we have rewritten their substrings into
shuffled terminals by \rrule3.  Finally rules \rrule4 and \rrule5
allow rewriting to occur at any position within a string. So in our
example grammar we have $\literal
T\rewrites\literal{Wpr}\rewrites\literal{mnpr}$ and $\literal
U\rewrites\literal{12345}$.  Since \literal{mnpr} and \literal{12345}
are both strings of terminals only, and since \literal{m12np3r45} is a
shuffle of \literal{mnpr} and \literal{12345}, we have
\begin{align*}
  \literal S &\rewrites \parterm{\literal S}{\literal T,\literal
    Y}\rewrites \parterm{\literal S}{\literal{Wpr},\literal
    Y}\rewrites \parterm{\literal
    S}{\literal{Wpr},\literal{12345}}\rewrites \parterm{\literal
    S}{\literal{mnpr},\literal{12345}}
  \\ & \rewrites\literal{m12np3r45}\enspace.
\end{align*}
So \literal{m12np3r45} is a string in our example grammar's language.

%
The figure also defines a rightmost marked rewriting relation which we
use to support the correctness of our parser.  Rightmost marked
rewriting adds a position marker $\posmark$ to the structure of
rewritten strings, and tracks rewrites to allow them only immediately
to the left of the marker.  Its first three rules are the same as for
unrestricted rewriting $\rewrites$, except for maintaining the marker
to the right of possible nonterminals, and to the left of guaranteed
terminals.  Moreover \mrrule5 and \rrule5 are also similar, since we
allow rightmost rewriting among any of the substrings to be shuffled.
In \mrrule4, marked rightmost rewriting is permissible only when there
are no nonterminals textually to the right of the one to be expanded.
We write $|u|$ to refer to the erasure of all position markers from
$u$.
\begin{lemma}
  \label{lemma:rewriting}
  \begin{enumerate}
  \item If \(u\rewrites^*u'\in\Sigma^*\), then \(u\posmark\Mrewrites^*\posmark u'\).
  \item If \(u_0\Mrewrites u_1\) and $|u_0|\neq|u_1|$, then \(|u_0|\rewrites|u_1|\).
  \end{enumerate}
\end{lemma}
The second clause is clear from erasing the marker, and from the more
lenient contexts of $\rewrites$.  For the first clause we can use an
additional intermediate relation which keeps the contextual
restrictions of $\Mrewrites$ but not the markers, to first argue for
the reordering of rewrites, and then the addition of markers.


\section{Generalized LR-shuffle parsing}
\label{sec:nd-glr-s}
Like the standard GLR parser, \parseralg\ defines an underlying
nondeterministic automaton, and optimizes a process for tracking all
possible traces through it.  We present the underlying
nondeterministic automaton in this section, and discuss
implementations in Section~\ref{sec:impl}.

Since we have extended the grammar and rewriting language for the
shuffle operator, we must revisit the notion of an \emph{item}. As in
the traditional case of rules from \(R\), the marker may be at the
beginning, the end, or between any two characters of the expansion.
We add two sorts of item to the classical notion.  Corresponding to a
rule in \(P\), we can have items $a\expTo\posmark\parterm
m{a_1,\cdots,a_n}$ and $a\expTo\parterm m{a_1,\cdots,a_n}\posmark$.
Note that these forms do not have the same syntax as for rewriting; in
each form, \(m\) is an integer at least as big as \(n\), rather than
the original nonterminal.  An \emph{initial} item has its marker at
the beginning of the right-hand side; a \emph{final} item, at the end.
Finally there is a placeholder item \(\indir a\), which we will use to
represent a transition to the subtask of recognizing a particular
shuffled substring.

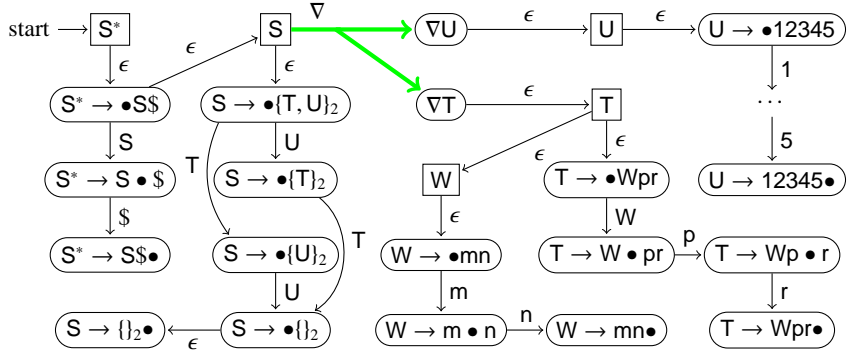
\begin{figure}[t]
  \begin{tikzpicture}[shorten >=1pt,node distance=10mm and 22mm,on grid,auto,
      station/.style = {draw, rectangle, inner sep=1mm},
      item/.style = {draw, rounded rectangle, inner sep=1mm},
      accept/.style = {draw, rounded rectangle={2pt}{rounded rectangle={2pt}}, inner sep=1mm}]
    \node[station,initial]  (Ss) {$\literal{S^*}$};
    \node[item] (Ss1) [below=of Ss]  {$\literal{S^*}\expTo\posmark\literal{S\$}$};
    \node[item] (Ss2) [below=of Ss1] {$\literal{S^*}\expTo\literal{S\posmark\$}$};
    \node[accept] (Ss3) [below=of Ss2] {$\literal{S^*}\expTo\literal{S\$}\posmark$};
    \node[station]  (S) [right=of Ss] {$\literal{S}$};
    \node[item]  (indU) [right=of S] {$\indir\literal{U}$};
    \node[item] (S1) [below=of S] {$\literal{S}\expTo\posmark\parterm 2{\literal T,\literal U}$};
    \node[item] (S2a) [below=of S1]   {$\literal{S}\expTo\posmark\parterm 2{\literal T}$};
    \node[item] (S2b) [below=of S2a]  {$\literal{S}\expTo\posmark\parterm 2{\literal U}$};
    \node[item] (S3) [below=of S2b]  {$\literal{S}\expTo\posmark\parterm 2{}$};
    \node[item] (S4) [left=of S3]  {$\literal{S}\expTo\parterm 2{}\posmark$};
    \node[item]  (indT) [right=of S1] {$\indir\literal{T}$};
    \node[station]  (T) [right=of indT]  {$\literal{T}$};
    \node[item] (T1) [below=of T]   {$\literal{T}\expTo\posmark\literal{Wpr}$};
    \node[item] (T2) [below=of T1]  {$\literal{T}\expTo\literal{W\posmark pr}$};
    \node[item] (T3) [right=of T2]  {$\literal{T}\expTo\literal{Wp\posmark r}$};
    \node[item] (T4) [below=of T3]  {$\literal{T}\expTo\literal{Wpr\posmark}$};
    \node[station]  (W) [below=of indT]  {$\literal{W}$};
    \node[item] (W1) [below=of W]   {$\literal{W}\expTo\posmark\literal{mn}$};
    \node[item] (W2) [below=of W1]  {$\literal{W}\expTo\literal{m\posmark n}$};
    \node[item] (W3) [right=of W2]  {$\literal{W}\expTo\literal{mn}\posmark$};
    \node[station]  (U) [right=of indU]  {$\literal{U}$};
    \node[item] (U1) [right=of U]   {$\literal{U}\expTo\posmark\literal{12345}$};
    \node (Udots) [below=of U1]  {$\cdots$};
    \node[item] (U6) [below=of Udots]  {$\literal{U}\expTo\literal{12345}\posmark$};

    \path[->] (Ss)  edge node {$\epsilon$} (Ss1);
    \path[->] (Ss1)  edge node {$\epsilon$} (S);
    \path[->] (Ss1) edge node {\literal{S}} (Ss2);
    \path[->] (Ss2) edge node {\literal{\$}} (Ss3);
    \path[->] (S)  edge node {$\epsilon$} (S1);
    \path[->] (S1) edge node {\literal{U}} (S2a);
    \path[->] (S1.south west) edge [bend right] node[above left] {\literal{T}} (S2b.north west);
    \path[->] (S2a.south east) edge [bend left=55] node[above right] {\literal{T}} (S3.north east);
    \path[->] (S2b) edge node {\literal{U}} (S3);
    \path[->] (S3) edge node {$\epsilon$} (S4);
    \path[->] (T)  edge node {$\epsilon$} (T1);
    \path[->] (T)  edge node {$\epsilon$} (W);
    \path[->] (T1) edge node {\literal{W}} (T2);
    \path[->] (T2) edge node {\literal{p}} (T3);
    \path[->] (T3) edge node {\literal{r}} (T4);
    \path[->] (W)  edge node {$\epsilon$} (W1);
    \path[->] (W1) edge node {\literal{m}} (W2);
    \path[->] (W2) edge node {\literal{n}} (W3);
    \path[->] (indU)  edge node {$\epsilon$} (U);
    \path[->] (indT)  edge node {$\epsilon$} (T);
    \path[->] (U)  edge node {$\epsilon$} (U1);
    \path[->] (U1) edge node {\literal{1}} (Udots);
    \path[->] (Udots) edge node {\literal{5}} (U6);
    \path[-] (S) edge [ultra thick, green] node {\textcolor{black}{$\indir$}} (3.1,0);
    \path[->] (3,0) edge [ultra thick, green] (indT);
    \path[->] (3,0) edge [ultra thick, green] (indU);
  \end{tikzpicture}
  \caption[Nondeterministic handle-finding automata for our example
    grammar.]{Nondeterministic handle-finding automata for our
    example grammar.  The hyperedge arising from the shuffle operator
    is highlighted in green.}
  \label{fig:handle-finding-nfa}
\end{figure}

\begin{figure}[t]
  \begin{tikzpicture}[shorten >=1pt,node distance=15mm and 22mm,on grid,auto,
      item/.style = {draw, rectangle, inner sep=1mm},
      accept/.style = {draw, rounded rectangle={2pt}{rounded rectangle={2pt}, inner sep=1mm}, inner sep=1mm}]
    \node[item,initial above]  (Ss) {
      \itemset{0}{$\literal{S^*}\expTo\posmark\literal{S\$}$
        \\ $\literal{S}\expTo\posmark\parterm 2{\literal T,\literal U}$}};
    \node[item] (S2a) [below left=of Ss]   {\itemset{1}{
        $\literal{S}\expTo\posmark\parterm 2{\literal T}$}};
    \node[item] (Ss2) [left=of Ss] {\itemset{16}{$\literal{S^*}\expTo\literal{S\posmark\$}$}};
    \node[item] (Ss3) [left=of Ss2] {\itemset{17}{$\literal{S^*}\expTo\literal{S\$}\posmark$}};
    \node[accept] (S2b) [below=of Ss]  {\itemset{2}{$\literal{S}\expTo\posmark\parterm 2{\literal U}$}};
    \node[item] (S3) [below=of S2a]  {\itemset{3}{$\literal{S}\expTo\posmark\parterm 2{}$
        \\ $\literal{S}\expTo\parterm 2{}\posmark$}};
    \node[item]  (U) [right=of S2b]  {\itemset{4}{$\indir\literal U$\\$\literal{U}\expTo\posmark\literal{12345}$}};
    \node (Udots) [below=of U]  {$\cdots$};
    \node[item] (U6) [left=of Udots]  {\itemset{9}{$\literal{U}\expTo\literal{12345}\posmark$}};
    \node[item]  (T) [above right=of U]  {
      \itemset{10}{
        $\indir\literal T$
        \\ $\literal{T}\expTo\posmark\literal{Wpr}$
        \\ $\literal{W}\expTo\posmark\literal{mn}$}
    };
    \node[item] (T2) [below=of T]  {\itemset{11}{$\literal{T}\expTo\literal{W\posmark pr}$}};
    \node[item] (T3) [below=of T2]  {\itemset{12}{$\literal{T}\expTo\literal{Wp\posmark r}$}};
    \node[item] (T4) [right=of T3]  {\itemset{13}{$\literal{T}\expTo\literal{Wpr\posmark}$}};
    \node[item] (W2) [right=of T]  {\itemset{14}{$\literal{W}\expTo\literal{m\posmark n}$}};
    \node[item] (W3) [below=of W2]  {\itemset{15}{$\literal{W}\expTo\literal{mn}\posmark$}};

    \path[->] (Ss) edge node {\literal{S}} (Ss2);
    \path[->] (Ss2) edge node {\literal{\$}} (Ss3);
    \path[->] (Ss) edge node {\literal{U}} (S2a);
    \path[->] (Ss) edge node {\literal{T}} (S2b);
    \path[->] (S2a) edge node {\literal{T}} (S3);
    \path[->] (S2b) edge node {\literal{U}} (S3);
    \path[->] (T) edge node {\literal{W}} (T2);
    \path[->] (T2) edge node {\literal{p}} (T3);
    \path[->] (T3) edge node {\literal{r}} (T4);
    \path[->] (T) edge node {\literal{m}} (W2);
    \path[->] (W2) edge node {\literal{n}} (W3);
    \path[->] (U)  edge node {\literal{1}} (Udots);
    \path[->] (Udots) edge node {\literal{5}} (U6);
    \path[-] (Ss) edge [ultra thick, green] node {\textcolor{black}{$\indir$}} (2.1,0);
    \path[->] (2,0) edge [ultra thick, green] (T);
    \path[->] (2,0) edge [ultra thick, green] (U);
  \end{tikzpicture}
  \caption[Conversion of the NDA of
    Figure~\protect\ref{fig:handle-finding-nfa} into a DFA.]{
    Conversion of the NDA of
    Figure~\protect\ref{fig:handle-finding-nfa} into a DFA.  Numbers
    in blue are an arbitrary assignment of a number to each state,
    which we use in Table~\protect\ref{table:tables}.}
  \label{fig:handle-finding-dfa}
\end{figure}
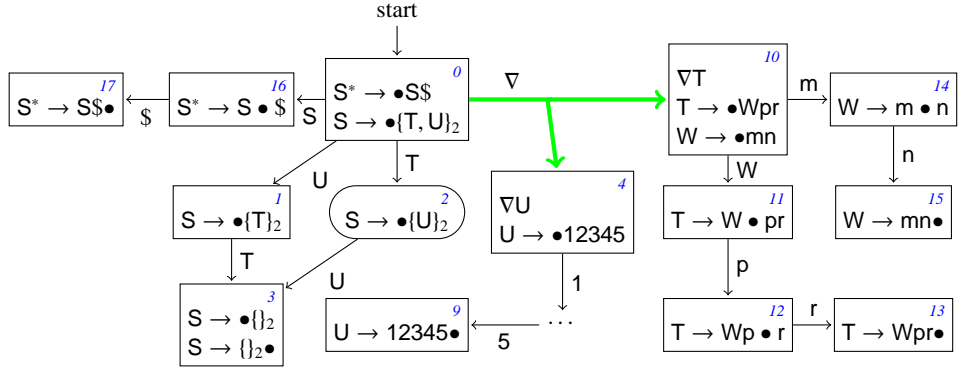

As usual when constructing an LR-style parser we use a handle-finding
deterministic finite automaton (DFA), derived from a nondeterministic
finite automaton (NFA) based on items as states.  In our motivating
application of plan recognition, lookahead to future actions to be
observed is not realistic, so we use an LR(0) handle-finding automaton
here, but we expect that for LR(1) or LALR(1) handle-finding we would
proceed similarly.  In addition to the standard transitions, and to
the usual \emph{stations} for the nonterminals
themselves~\cite{grune-jacobs-parsing:2008}, from an item
$a\expTo\posmark\parterm{n}{a_1,\ldots,a_n}$ we have a transition
labelled $a_i$ to
$a\expTo\posmark\parterm{n}{a_1,\ldots,a_{i-1},a_{i+1},\ldots,a_n}$
for each $i$, plus an $\epsilon$-transition from any
$a\expTo\posmark\parterm{n}{}$ to $a\expTo\parterm{n}{}\posmark$.  We
annotate the NDA for our handle finder with
\emph{hyperedges} linking from one state to several.  Specifically the
graph of states and edges of our extension forms an arc-labeled
F-directed hypergraph~\cite{gallo-et-al:hypergraphs:1993}.
Of course the
hyperedges have no impact on the operation of an NDA;
one does not ``run'' the handle-finding automaton in any real sense.
We use the hyperedges for bookkeeping, translating
them to the DFA and generating particular action table entries based on them.
Where there is such a hyperedge edge from an item $I$ to items $I_i$ in the
NDA, we expect that any state containing $I$ in the DFA would have a
similar hyperedge to the least sets containing the $I_i$.
From an
item $a\expTo \posmark \parterm n{a_1,\ldots,a_n}$ we add a hyperedge
labelled $\indir$, a symbol not in the original grammar, to $n$
indirection items $\indir a_1,\cdots,\indir a_n$; from each
indirection item $\indir a_i$ we have an $\epsilon$-transition to the
station for $a_i$.  By convention we designate a nonterminal
\(\literal{S^*}\) and terminal \(\eosmark\) which are not in the
original grammar, and take the initial state of the NFA to be the
station for \(\literal{S^*}\).  Figure~\ref{fig:handle-finding-nfa}
shows the nondeterministic handle-finding automaton for our example
grammar, and Figure~\ref{fig:handle-finding-dfa} shows its translation
to a DFA.

In non-generalized parsers, we expect that each goto/action table
position will have exactly one entry (which might be to fail).  This
allows the creation of a \emph{deterministic} parser;
otherwise the presence of a \emph{defective state} with multiple
conflicting entries makes it unclear which of the steps should be
followed.  The essential insight of Tomita's GLR parser
is that for most grammars, and in particular those which
are of practical use, we can efficiently track \emph{all} of the
possible actions.  So we do not insist that a \parseralg\ goto/action
table be free of conflicts.  As usual the table is indexed by state
and by symbol; each table entry \(\gotoaction{\stateS}{a}\) includes a
set of instructions, which are empty except as described by these
rules:
\begin{enumerate}
\item For each \(a\expTo u\posmark\in\stateS\), add \(\reduce(a\expTo
      u\posmark)\) to entry \(\gotoaction{\stateS}{a'}\) for all \(a'\).
\item For each simple edge \(\stateS\transby a\stateS'\), add
  \(\shift(\stateS')\) to entry \(\gotoaction{\stateS}a\).
\item For each hyperedge
  \(\stateS\transby{\indir}\stateS_1,\cdots,\stateS_n\) and simple
  edge \(\stateS_i\transby s\stateS'_i\) where $s\in\Sigma$, include
  \(\shift(\stateS_i\rightarrow\stateS'_i;\stateS_1,\cdots,\stateS_{i-1},\stateS_{i+1},\stateS_n)\)
  to entry \(\gotoaction{\stateS_0}s\).
\item For any transition by $\literal \$$ to a state containing the
  accepting item,
  $\stateS\transby{\literal\$}\{\literal{S^*}\expTo\literal{S\$}\posmark,\ldots\}$,
  we add $\accept$ to $\gotoaction{\stateS}{\literal\$}$
\end{enumerate}
The rules for \(\reduce\), for $\shift$ from a simple edge, and for
$\accept$ are exactly as for classical CFG parsing.  The third rule
addresses hyperedges.  Through the hyperedges we
identify the initial states for recognizing the shuffled strings, but
a hyperedge does not correspond to recognizing an actual piece of
the input.  So the shift operations which initiate recognizing a
shuffle derive from two edges, the hyperedge plus a
single subsequent edge which actually consumes a piece of input.
Table~\ref{table:tables} shows the first few rows and columns of the
table for our running example.  Note that our restriction to $P$
against the referencing further rules in $P$ simplifies Rule~3 here,
otherwise we must chase through, and accumulate concurrent terms from,
several rules.
\begin{table}[t]
  \[\begin{array}{c|c@{~~~~}c@{~~~~}c@{~~~~}c@{~~~~}c@{~~~~}c@{~~~~}c}
  \multicolumn{1}{c}{}
  & \literal S & \literal T & \literal U & \literal W & \literal m &
  \literal 1 & \cdots
  \\ \cline{2-7}
     0 & \shift(16) & \shift 7 & \shift 1 & - &
  \shift(10\rightarrow14;4) & \shift(4\rightarrow5;10)
  \\ 1 & - & \shift 3 & - & - & - & -
  \\ 2 & - & - & \shift 3 & - & - & -
  \\ 3 & \multicolumn{6}{c}{\dotfill\reduce{S\expTo\parterm2{}\posmark}\dotfill}
  \\ 4 & - & - & - & - & - & \shift{5}
  \\ \multicolumn{1}{c}{\vdots}
  \end{array}\]
  \caption[Part of the goto/action table for our example
    grammar.]{Part of the goto/action table for our example grammar.
    In Row 3, the reduce action appears in every column.}
  \label{table:tables}
\end{table}

\begin{figure}[t]
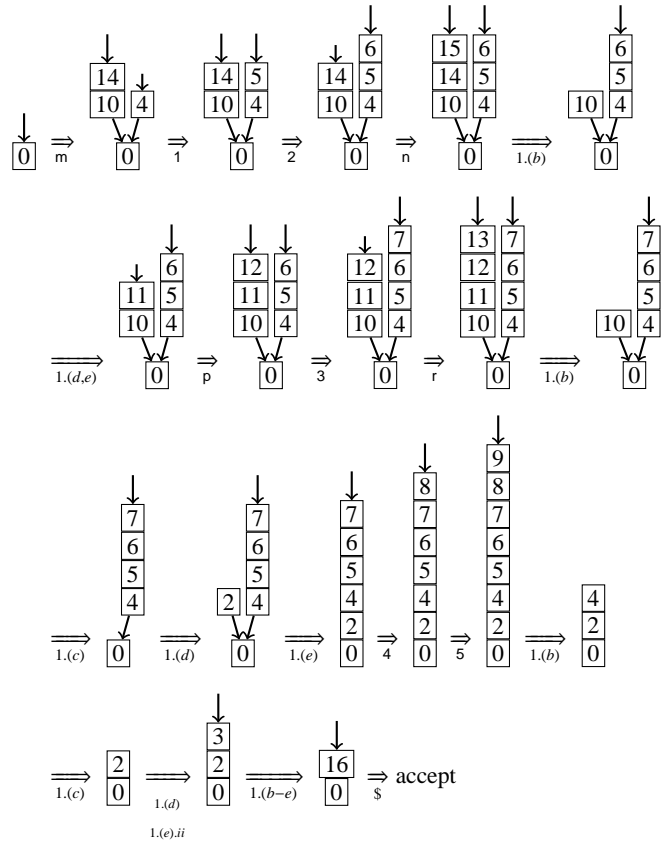

  \begin{align*}
    \cactus{
      \node[arrowbase](b1){}; \\[3mm]
      \node(zero){0}; \\
    }{\draw[->,black,thick] (b1.north) -- (zero); }
    &
    \stepStack{\literal m}
    \cactus{
      \node[arrowbase](b1){}; \\[3mm]
      \node(t1){14}; \& \& \node[arrowbase](b2){}; \\
      \node(ten){10}; \& \& \node(four){4}; \\[1em]
      \& \node(zero){0}; \& \\
    }{\draw[->,black,thick] (b1.north) -- (t1);
      \draw[->,black,thick] (b2.north) -- (four);
      \draw[->,black,thick] (ten) -- (zero);
      \draw[->,black,thick] (four) -- (zero);}
    \stepStack{\literal 1}
    \cactus{
      \node[arrowbase](b1){}; \& \& \node[arrowbase](b2){}; \\[3mm]
      \node(t1){14}; \& \& \node(t2){5}; \\
      \node(ten){10}; \& \& \node(four){4}; \\[1em]
      \& \node(zero){0}; \& \\
    }{\draw[->,black,thick] (b1.north) -- (t1);
      \draw[->,black,thick] (b2.north) -- (t2);
      \draw[->,black,thick] (ten) -- (zero);
      \draw[->,black,thick] (four) -- (zero);}
    \stepStack{\literal 2}
    \cactus{
       \& \& \node[arrowbase](b2){}; \\[3mm]
      \node[arrowbase](b1){}; \& \& \node(t2){6}; \\
      \node(t1){14}; \& \& \node{5}; \\
      \node(ten){10}; \& \& \node(four){4}; \\[1em]
      \& \node(zero){0}; \& \\
    }{\draw[->,black,thick] (b1.north) -- (t1);
      \draw[->,black,thick] (b2.north) -- (t2);
      \draw[->,black,thick] (ten) -- (zero);
      \draw[->,black,thick] (four) -- (zero);}
    \stepStack{\literal n}
    \cactus{
      \node[arrowbase](b1){}; \& \& \node[arrowbase](b2){}; \\[3mm]
      \node(t1){15}; \& \& \node(t2){6}; \\
      \node{14}; \& \& \node{5}; \\
      \node(ten){10}; \& \& \node(four){4}; \\[1em]
      \& \node(zero){0}; \& \\
    }{\draw[->,black,thick] (b1.north) -- (t1);
      \draw[->,black,thick] (b2.north) -- (t2);
      \draw[->,black,thick] (ten) -- (zero);
      \draw[->,black,thick] (four) -- (zero);}
    \stepStack{1.(b)}
    \cactus{
       \& \& \node[arrowbase](b2){}; \\[3mm]
       \& \& \node(t2){6}; \\
       \& \& \node{5}; \\
      \node(ten){10}; \& \& \node(four){4}; \\[1em]
      \& \node(zero){0}; \& \\
    }{\draw[->,black,thick] (b2.north) -- (t2);
      \draw[->,black,thick] (ten) -- (zero);
      \draw[->,black,thick] (four) -- (zero);}
    \\ & 
    \stepStack{1.(d,e)}
    \cactus{
       \& \& \node[arrowbase](b2){}; \\[3mm]
      \node[arrowbase](b1){}; \& \& \node(t2){6}; \\
      \node(t1){11}; \& \& \node{5}; \\
      \node(ten){10}; \& \& \node(four){4}; \\[1em]
      \& \node(zero){0}; \& \\
    }{\draw[->,black,thick] (b1.north) -- (t1);
      \draw[->,black,thick] (b2.north) -- (t2);
      \draw[->,black,thick] (ten) -- (zero);
      \draw[->,black,thick] (four) -- (zero);}
    \stepStack{\literal p}
    \cactus{
      \node[arrowbase](b1){}; \& \& \node[arrowbase](b2){}; \\[3mm]
      \node(t1){12}; \& \& \node(t2){6}; \\
      \node{11}; \& \& \node{5}; \\
      \node(ten){10}; \& \& \node(four){4}; \\[1em]
      \& \node(zero){0}; \& \\
    }{\draw[->,black,thick] (b1.north) -- (t1);
      \draw[->,black,thick] (b2.north) -- (t2);
      \draw[->,black,thick] (ten) -- (zero);
      \draw[->,black,thick] (four) -- (zero);}
    \stepStack{\literal 3}
    \cactus{
      \& \& \node[arrowbase](b2){}; \\[3mm]
      \node[arrowbase](b1){}; \& \& \node(t2){7}; \\
      \node(t1){12}; \& \& \node{6}; \\
      \node{11}; \& \& \node{5}; \\
      \node(ten){10}; \& \& \node(four){4}; \\[1em]
      \& \node(zero){0}; \& \\
    }{\draw[->,black,thick] (b1.north) -- (t1);
      \draw[->,black,thick] (b2.north) -- (t2);
      \draw[->,black,thick] (ten) -- (zero);
      \draw[->,black,thick] (four) -- (zero);}
    \stepStack{\literal r}
    \cactus{
      \node[arrowbase](b1){}; \& \& \node[arrowbase](b2){}; \\[3mm]
      \node(t1){13}; \& \& \node{7}; \\
      \node{12}; \& \& \node{6}; \\
      \node{11}; \& \& \node{5}; \\
      \node(ten){10}; \& \& \node(four){4}; \\[1em]
      \& \node(zero){0}; \& \\
    }{\draw[->,black,thick] (b1.north) -- (t1);
      \draw[->,black,thick] (b2.north) -- (t2);
      \draw[->,black,thick] (ten) -- (zero);
      \draw[->,black,thick] (four) -- (zero);}
    \stepStack{1.(b)}
    \cactus{
       \& \& \node[arrowbase](b2){}; \\[3mm]
       \& \& \node(t2){7}; \\
       \& \& \node{6}; \\
       \& \& \node{5}; \\
      \node(ten){10}; \& \& \node(four){4}; \\[1em]
      \& \node(zero){0}; \& \\
    }{\draw[->,black,thick] (b2.north) -- (t2);
      \draw[->,black,thick] (ten) -- (zero);
      \draw[->,black,thick] (four) -- (zero);}
    \\ &
    \stepStack{1.(c)}
    \cactus{
      \& \node[arrowbase](b1){}; \\[3mm]
      \& \node(t1){7}; \\
      \& \node{6}; \\
      \& \node{5}; \\
      \& \node(four){4}; \\[1em]
      \node(zero){0}; \& \\
    }{\draw[->,black,thick] (b1.north) -- (t1);
       \draw[->,black,thick] (four) -- (zero);}
    \stepStack{1.(d)}
    \cactus{
      \& \& \node[arrowbase](b1){}; \\[3mm]
      \& \& \node(t1){7}; \\
      \& \& \node{6}; \\
      \& \& \node{5}; \\
      \node(two){2}; \& \& \node(four){4}; \\[1em]
      \& \node(zero){0}; \& \\
    }{\draw[->,black,thick] (b1.north) -- (t1);
      \draw[->,black,thick] (two) -- (zero);
      \draw[->,black,thick] (four) -- (zero);}
    \stepStack{1.(e)}
    \cactus{
      \node[arrowbase](b1){}; \\[3mm]
      \node(t1){7}; \\
      \node{6}; \\
      \node{5}; \\
      \node{4}; \\
      \node{2}; \\
      \node{0}; \\
    }{\draw[->,black,thick] (b1.north) -- (t1);
      }
    \stepStack{\literal 4}
    \cactus{
      \node[arrowbase](b1){}; \\[3mm]
      \node(t1){8}; \\
      \node{7}; \\
      \node{6}; \\
      \node{5}; \\
      \node{4}; \\
      \node{2}; \\
      \node{0}; \\
    }{\draw[->,black,thick] (b1.north) -- (t1);
      }
    \stepStack{\literal 5}
    \cactus{
      \node[arrowbase](b1){}; \\[3mm]
      \node(t1){9}; \\
      \node{8}; \\
      \node{7}; \\
      \node{6}; \\
      \node{5}; \\
      \node{4}; \\
      \node{2}; \\
      \node{0}; \\
    }{\draw[->,black,thick] (b1.north) -- (t1);
      }
    \stepStack{1.(b)}
    \cactus{
      \node(t1){4}; \\
      \node{2}; \\
      \node{0}; \\
    }{}
    \\ &
    \stepStack{1.(c)}
    \cactus{
      \node(t1){2}; \\
      \node{0}; \\
    }{}
    \stepStack{\begin{array}{@{}c@{}}\scriptscriptstyle
        1.(d)\\\scriptscriptstyle 1.(e).ii\end{array}}
    \cactus{
      \node[arrowbase](b1){}; \\[3mm]
      \node(t1){3}; \\
      \node{2}; \\
      \node{0}; \\
    }{\draw[->,black,thick] (b1.north) -- (t1);
      }
    \stepStack{1.(b-e)}
    \cactus{
      \node[arrowbase](b1){}; \\[3mm]
      \node(t1){16}; \\
      \node{0}; \\
    }{\draw[->,black,thick] (b1.north) -- (t1);
      }
    \stepStack{\literal\$}
    \accept
  \end{align*}
  \caption{Parser actions for the string $\literal{m12np3r45}$ with
    our running example.}
  \label{fig:parse-example}
\end{figure}

With the goto/action table in hand, we can specify a nondeterministic
automaton which recognizes strings in our \grammarclass.
Figure~\ref{fig:alg} presents our algorithm.  The runtime state of the
parser is a cactus stack --- a tree which grows and shrinks at the
leaves --- of states from the handle finding DFA.  For clarity we
explicitly mark the nodes which we consider to be stack tops, since
we detail the creation of nodes which are only temporarily at a leaf
in the tree.  We write $\beta:\stateS$ to name a stack top $\beta$
containing the state $\stateS$.  Figure~\ref{fig:parse-example} shows
the manipulations to the cactus stack as \parseralg\ recognizes the
string \literal{m12np3r45} in our example grammar.

Step~1(c) of the algorithm shows the purpose of the $\indir\literal a$
nodes.  The reduce operations in LR parsers rely on the number of
states pushed onto the stack for the right-hand side of a rule being
the same as the length of that right-hand side.  But when we split the
stack for shuffled substrings we start the substacks with an initial
state that does not correspond to any recognized symbol.  The presence
of an indirection to $\indir\literal{a}$ in an item set rectifies this
offset.  But since the $\indir\literal{a}$ item would not appear in
the state corresponding to a sequential use of \literal{a} --- which
would follow an $\epsilon$-link to \literal{a}'s station instead of to
$\indir\literal{a}$ --- we will not disrupt stack operations in
non-shuffled cases.

\begin{figure}[t]
  \begin{algorithm}[\parseralg\ parsing]
    \label{alg}
  Initially the single state on the stack is the DFA's initial state.
  For each symbol $s$ of the input string including the end-of-string
  marker $\literal\$$:

  \begin{enumerate}
  \item For each reducible stack top \(\alpha:\stateS\), if
    \(\reduce{a\expTo u\posmark}\in\stateS\), then we may choose to
    reduce that rule:
    
    \begin{enumerate}
    \item Drop \(\alpha\) as a stack top.

    \item Pop \(|u|\) nodes from \(\alpha\) to node \(\alpha'\).

    \item If \(\alpha':\stateS'\) contains the indirection node for
      $a$, \(\indir a\in\stateS'\), pop one additional time to
      \(\alpha''\), else take \(\alpha''\) to be just \(\alpha'\).
      
    \item Let \(\alpha'':\stateS_0\), and choose some shift operation
      \(\shift(\stateS'_0)\in\gotoaction{\stateS_0}a\), or raise an
      error if there is no possible shift.  Create \(\beta:\stateS'_0\)
      with parent \(\alpha''\).

    \item If $\alpha''$ has other child nodes:

      \begin{enumerate}
      \item Then update the other children of \(\alpha''\) to have
        \(\beta\) as their parent.

      \item Else take $\beta$ as a stack top.
      \end{enumerate}
    \end{enumerate}

  \item Choose a stack top \(\alpha:\stateS\) with a stack or accept
    operation in \(\gotoaction{\stateS}s\) (or reject if there is no
    such \(\alpha\)), and choose one of those operations.
    \begin{enumerate}
    \item If the operation is $\accept$, then the parser accepts the
      string.
    \item If the operation is $\shift(\stateS)$, then create
      \(\beta:\stateS'\) with parent \(\alpha\), and replace
      \(\alpha\) as a stack top with \(\beta\).
    \item If the operation is
      $\shift(\stateS_0\rightarrow\stateS'_0;\stateS_1,\cdots,\stateS_n)\in$,
      then create \(\beta_i:\stateS_i\) with parent \(\alpha\) for
      each \(0\leq i\leq n\), and moreover create
      \(\beta'_0:\stateS'_0\) with parent $\beta_0$. Replace
      \(\alpha\) as a stack top with $\beta'_0$ and the
      \(\beta_1,\cdots,\beta_n\).
    \end{enumerate}
  \end{enumerate}
  After the end-of-string marker $\literal\$$ if we have
  not accepted the input, then we reject it.
\end{algorithm}
  \caption{The nondeterministic \parseralg\ parsing algorithm.}
  \label{fig:alg}
\end{figure}

\begin{lemma}[Main]
  \label{lemma:correctness}
  Algorithm~\ref{alg} accepts a string $u\in\Sigma$ under a grammar
  with starting symbol $S$ if and only if
  $S\posmark\Mrewrites^*\posmark u$.
\end{lemma}
  Specifically, the manipulations to the cactus stack under the
  \parseralg\ algorithm correspond to the reverse of an $\Mrewrites$
  sequence:
  \begin{itemize}
  \item Application of a $\reduce$ in Step~1 of the algorithm
    corresponds to a rewrite by Rule \mrrule1.
  \item Application of the simple shift operation in Step~2.(b)
    corresponds to a rewrite by rule \mrrule6.
  \item Application of the shuffle-decomposing shift operation of
    Step~2.(c) corresponds to a rewrite by Rule~\mrrule2, followed by
    a rewrite by Rule~\mrrule3.
  \end{itemize}

Lemmas~\ref{lemma:rewriting} and~\ref{lemma:correctness} justify the
correctness of the \parseralg\ parser.
\begin{theorem}[Correctness]
  Algorithm~\ref{alg} accepts a string $u\in\Sigma^*$ if and only if
  $S\rewrites^*u$.
\end{theorem}
  
\section{Necessary optimizations}
\label{sec:impl}

An important aspect of GLR is the construction of a single structure
to represent all current possible parse stacks.  To make the
representation reasonable, stack structure is shared whenever
possible.  The obvious case is that when two different actions are
possible for some stack top, we can simply branch the stack, so that
the common stack bottom is shared.  Less obvious, but just as
important, is the idea that common stack tops should be shared as
well: when a node for some state $\stateS$ would be pushed onto $n$
different stacks for one input symbol, we in fact allocate only one
single node, with pointers to all of the previous tops-of-stacks.
This one single node is taken as the top of a stack, as opposed to
maintaining $n$ different stacks.  Later, a reduce action can pop
nodes past the point where the several stacks join, resulting in
several stacks.  Tomita dubbed this construction the
\emph{graph-structured stack}~\cite{tomita-ijcai-1985}.

\begin{figure}[t]
  \begin{center}
    (a)\small
    \begin{tikzpicture}[shorten >=1pt,node distance=15mm and 30mm,on grid,auto,
        item/.style = {draw, rectangle, inner sep=1mm},
        accept/.style = {draw, rounded rectangle={2pt}{rounded rectangle={2pt}, inner sep=1mm}, inner sep=1mm}]
      \node[item]  (S) {\itemset{20}{$\literal{S}\expTo\posmark\parterm 3{\literal{T,U,V}}$}};
      \node[item]  (U) [right=of S]  {\itemset{40}{$\indir\literal U$\\$\literal{U}\expTo\posmark\literal{abc}$}};
      \node[item] (U1) [right=of U]  {\itemset{41}{$\literal{U}\expTo\literal{a\posmark bc}$}};
      \node[item] (U2) [right=of U1] {\itemset{42}{$\literal{U}\expTo\literal{ab\posmark c}$}};
      \node[item]  (T) [above=of U]  {\itemset{30}{$\indir\literal T$\\$\literal{T}\expTo\posmark\literal{ab}$}};
      \node[item] (T1) [right=of T]  {\itemset{31}{$\literal{T}\expTo\literal{a\posmark b}$}};
      \node[item] (T2) [right=of T1] {\itemset{32}{$\literal{T}\expTo\literal{ab\posmark}$}};
      \node[item]  (V) [below=of U]  {\itemset{50}{$\indir\literal V$\\$\literal{V}\expTo\posmark\literal{cc}$}};
      
      \path[->] (T) edge node {\literal{a}} (T1);
      \path[->] (U) edge node {\literal{a}} (U1);
      \path[->] (T1) edge node {\literal{b}} (T2);
      \path[->] (U1) edge node {\literal{b}} (U2);
      \path[-]  (S) edge [ultra thick, green] node {\textcolor{black}{$\indir$}} (1.6,0);
      \path[->] (1.6,0) edge [ultra thick, green] (T.west);
      \path[->] (1.6,0) edge [ultra thick, green] (U);
      \path[->] (1.6,0) edge [ultra thick, green] (V.west);
    \end{tikzpicture}
    \\[1em]
    (b)
    \begin{tabular}[b]{ll}
      & $\Nearrow_{\literal a}$
      \wcactus{
        \node[arrowbase](b1){}; \\[3mm]
        \node(s31){31}; \&\& \node[arrowbase](b2){}; \&\& \node[arrowbase](b3){}; \\
        \node(s30){30}; \&\& \node(s40){40}; \&\& \node(s50){50}; \\[1em]
        \&\& \node(s20){20}; \&\& \\
      }{\draw[->,black,thick] (b1.north) -- (s31);
        \draw[->,black,thick] (b2.north) -- (s40);
        \draw[->,black,thick] (b3.north) -- (s50);
        \draw[->,black,thick] (s30) -- (s20);
        \draw[->,black,thick] (s40) -- (s20);
        \draw[->,black,thick] (s50) -- (s20);}
      $\stepStack{\literal b}  $
      \cactus{
        \node[arrowbase](b2){}; \&\& \node[arrowbase](b3){}; \\[3mm]
        \node(s40){40}; \&\& \node(s50){50}; \\[1em]
        \&\node(s21){21}; \\
        \&\node(s20){20}; \\
      }{\draw[->,black,thick] (b2.north) -- (s40);
        \draw[->,black,thick] (b3.north) -- (s50);
        \draw[->,black,thick] (s40) -- (s21);
        \draw[->,black,thick] (s50) -- (s21);}
      \\[-6mm] \wcactus{
        \node[arrowbase](b1){}; \&\& \node[arrowbase](b2){}; \&\& \node[arrowbase](b3){}; \\[3mm]
        \node(s30){30}; \&\& \node(s40){40}; \&\& \node(s50){50}; \\[1em]
        \&\& \node(s20){20}; \&\& \\
      }{\draw[->,black,thick] (b1.north) -- (s30);
        \draw[->,black,thick] (b2.north) -- (s40);
        \draw[->,black,thick] (b3.north) -- (s50);
        \draw[->,black,thick] (s30) -- (s20);
        \draw[->,black,thick] (s40) -- (s20);
        \draw[->,black,thick] (s50) -- (s20);}
      \\[-9mm]
      & \raisebox{1cm}{${}_{\literal a}\!\!\Searrow$}
      \wcactus{
        \&\& \node[arrowbase](b2){}; \\[3mm]
        \node[arrowbase](b1){}; \&\& \node(s41){41}; \&\& \node[arrowbase](b3){}; \\
        \node(s30){30}; \&\& \node(s40){40}; \&\& \node(s50){50}; \\[1em]
        \&\& \node(s20){20}; \&\& \\
      }{\draw[->,black,thick] (b1.north) -- (s30);
        \draw[->,black,thick] (b2.north) -- (s41);
        \draw[->,black,thick] (b3.north) -- (s50);
        \draw[->,black,thick] (s30) -- (s20);
        \draw[->,black,thick] (s40) -- (s20);
        \draw[->,black,thick] (s50) -- (s20);}
      $\stepStack{\literal b}$
      \wcactus{
        \&\& \node[arrowbase](b2){}; \\[3mm]
        \&\& \node(s42){42}; \&\& \\
        \node[arrowbase](b1){}; \&\& \node(s41){41}; \&\& \node[arrowbase](b3){}; \\
        \node(s30){30}; \&\& \node(s40){40}; \&\& \node(s50){50}; \\[1em]
        \&\& \node(s20){20}; \&\& \\
      }{\draw[->,black,thick] (b1.north) -- (s30);
        \draw[->,black,thick] (b2.north) -- (s42);
        \draw[->,black,thick] (b3.north) -- (s50);
        \draw[->,black,thick] (s30) -- (s20);
        \draw[->,black,thick] (s40) -- (s20);
        \draw[->,black,thick] (s50) -- (s20);}
    \end{tabular}
    \hspace*{\fill}
    (c)\wcactus{
      \& \node[arrowbase](b1){}; \& \& \node[arrowbase](b2){}; \\[3mm]
      \node[arrowbase](g1){}; \& \node(s31){31};  \&\node[arrowbase](g2){};\& \node(s41){41}; \&
      \node[arrowbase](g3){}; \& \node[arrowbase](b3){}; \\
      \& \node(s30){30}; \&\& \node(s40){40}; \&\& \node(s50){50}; \\[1em]
      \& \&\& \node(s20){20}; \&\& \\
    }{\draw[->,red,thick] (b1.north) -- (s31);
      \draw[->,green,thick] (b2.north) -- (s41);
      \draw[->,green,thick] (b3.north) -- (s50);
      \draw[->,green,thick] (g1.north) -- (s30.north west);
      \draw[->,red,thick] (g2.north) -- (s40.north west);
      \draw[->,red,thick] (g3.north) -- (s50);
      \draw[->,black,thick] (s30) -- (s20);
      \draw[->,black,thick] (s40) -- (s20);
      \draw[->,black,thick] (s50) -- (s20);}
    \\[1em]
    (d)\raisebox{8mm}{
      \textcolor{blue}{$\alpha_-$}\hspace{1mm}$\stepStack{\literal a}\hspace{-3mm}$
      \begin{tikzpicture}[
          baseline={([yshift=-.8ex]current bounding box.center)},
          every node/.style={green,inner sep=3pt}
        ]
        \pgfsetmatrixcolumnsep{0mm}
        \pgfsetmatrixrowsep{3mm}
        \pgfmatrix{rectangle}{center}{mymatrix}
                  {\pgfusepath{}}{\pgfpointorigin}{\let\&=\pgfmatrixnextcell}
                  {
                    \& \& \node(K){$\kpull K\!:\!{\alpha,-}\kpull$}; \\
                    \& \& \node(or){$\vee$}; \\[-2mm]
                    \& \node(and1){$\wedge$}; \& \& \node(and2){$\wedge$}; \\[1mm]
                    \node(gamma){$\gamma_K$}; \& \node(zeta){$\zeta_K$}; \& \node(delta){$\delta_K$}; \& \node(epsilon){$\epsilon_K$}; \& \node(eta){$\eta_K$}; \\
                  }
                  \draw[->,green] (K) -- (or);
                  \draw[->,green] (or) -- (and1);
                  \draw[->,green] (or) -- (and2);
                  \draw[->,green] (and1) -- (gamma);
                  \draw[->,green] (and1) -- (zeta);
                  \draw[->,green] (and1) -- (eta.north west);
                  \draw[->,green] (and2) -- (delta);
                  \draw[->,green] (and2) -- (epsilon);
                  \draw[->,green] (and2) -- (eta);
      \end{tikzpicture}
      $\hspace{-3mm}\stepStack{\literal b}\hspace{-3mm}$
      \begin{tikzpicture}[
          baseline={([yshift=-.8ex]current bounding box.center)},
          every node/.style={red,inner sep=3pt}
        ]
        \pgfsetmatrixcolumnsep{0mm}
        \pgfsetmatrixrowsep{3mm}
        \pgfmatrix{rectangle}{center}{mymatrix}
                  {\pgfusepath{}}{\pgfpointorigin}{\let\&=\pgfmatrixnextcell}
                  {
                    \&\& \node(or){$\vee$}; \\[0mm]
                    \& \node(and1){$\kpull K\!:\!{\alpha,-}\kpull$}; \&\& \node(and2){$\kpull K\!:\!{\beta,-}\kpull$}; \\[1mm]
                    \node(gamma){$\gamma_K$}; \& \node(mu){$\mu_K$}; \&\& \node(epsilon){$\epsilon_K$}; \& \node(eta){$\eta_K$}; \\
                  }
                  \draw[->,red] (or) -- (and1);
                  \draw[->,red] (or) -- (and2);
                  \draw[->,red] (and1) -- (gamma);
                  \draw[->,red] (and1) -- (mu);
                  \draw[->,red] (and1) -- (eta.north west);
                  \draw[->,red] (and2) -- (epsilon);
                  \draw[->,red] (and2) -- (eta);
      \end{tikzpicture}
      \hspace{\fill}
      \wwcactus{
        \&  \& \node[red,label={[red]left:$\mu\!:$}](mu){42}; \\
        \node[red,label={[red]left:$\beta\!:$}](beta){21};
        \& \node[green,label={[green]left:$\delta\!:$}](delta){31};
        \&\node[green,label={[green]left:$\zeta\!:$}](zeta){41}; \\
        \node[blue,label={[blue]left:$\alpha\!:$}](alpha){20};
        \& \node[green,label={[green]left:$\delta\!:$}](gamma){30};
        \& \node[green,label={[green]left:$\epsilon\!:$}](epsilon){40};
        \& \node[green,label={[green]left:$\eta\!:$}](eta){50}; \\
      }{}
    }
  \end{center}
  \caption{Example grammar elements and cactus stacks.}
  \label{fig:motivateUpper}
\end{figure}
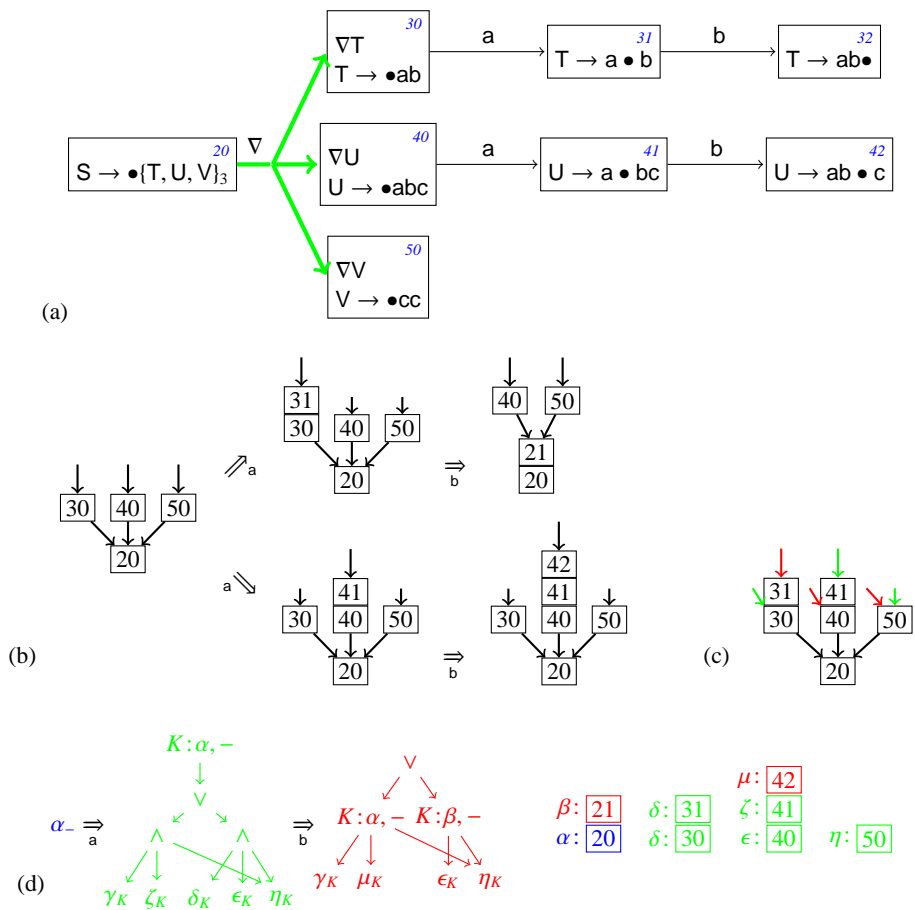

We adopt the graph-structured stack for \parseralg, but the midstack
mutation performed at Step~1.(e).i of the algorithm complicates its
use.  In the nondeterministic parser it is simple enough to mutate the
middle of a stack, but not all of the stacks superposed in
graph-structuring will undergo the same mutations.  Consider the
fragment of DFA in Figure~\ref{fig:motivateUpper}(a).  The string
\literal{ab} could be attributed to either \literal{T} or \literal{U},
and we have two possible parses, shown in
Figure~\ref{fig:motivateUpper}(b).  We can easily imagine superposing
the stacks resulting after recognizing just \literal{a}:
Figure~\ref{fig:motivateUpper}(c) shows the combined stacks, with two
sets of stack tops depicted in different colors.  But the stacks
resulting after \literal{b} are more difficult to combine, since they
differ not only at their tops but also internally.

To store stacks in such cases we use not one but two graphs, which we
distinguish as \emph{upper} and \emph{lower}.  The lower graph holds
parser states as in Tomita's graph-structured stack, but disconnected
at the junctions arising from shuffle operations.  The relationship
between these graph fragments, as well as the sets of stack tops, are
maintained in the upper graph.  The upper graph has the structure of
an and/or tree, possibly with shared substructure.  Or-nodes reflect
different possible parses; and-nodes organize recognition of shuffled
substrings.  To chain together substacks in the lower graph, we bind
\emph{controllers} in and-nodes, and reference both a stack top and
one of these controllers in each leaf node.  The binding of a
controller in an and-node associates the controller with both a prior
stack top, and a parent controller.  So essentially the controllers
form a linked tree of stacks from the lower graph, which taken
together assemble the cactus stacks of the nondeterministic algorithm.
Separating the binding of a controller to a parent stack top on the
one hand, from the
stack tops in leaf nodes on the other hand, allows us to locally
mutate controller bindings in a way that restricts the scope of the
effect.

We illustrate the use of controllers in
Figure~\ref{fig:motivateUpper}(d), with the series of upper graphs to
the left, and the lower graph on the right.  Note that we label each
lower graph node with a Greek letter, and that lower graph nodes take
the same color as the upper graph from which they were added.  Before
processing input symbol \literal a, we have a single stack top
$\alpha$ and a void controller $-$.  Starting to parse the shuffled
substring, \parseralg\ creates the controller $K$, associating it with
$\alpha$ as the previous top of stack.  Since there are two ways to
understand this \literal{a}, as either the first symbol of \literal{T}
or as the first symbol of \literal{U}, we see below the new binding
two different groups of stack tops.  The or node labelled $\vee$ sits
below the binding to spare the work of duplicating it; the and-nodes
labelled $\wedge$, which do not bind new controllers, each identify a
set of three stack tops reflecting the state of parsing the shuffled
strings.  When processing the \literal{b} we find one case where
reducing \literal{T} requires mutating the middle of the stack by
pushing a new node above $\alpha$, and one case where we do not.
Reconciling the two does require separate bindings for $K$, and the
resulting pair appears under an or-node.  The meaning of a controller
reference in a tree leaf is determined with respect to a particular
path from the root of the upper graph through various and-nodes with
bindings and to the leaf.  So the leaves do not change as the upper
graph evolves unless a new node is pushed onto them.

For each input symbol we traverse the upper graph, constructing a new
upper graph while reusing as much of the previous graph as possible.
When traversing an or-node we discard any children for which there is
no way to advance with the next input.  For an and-node we require
exactly one of its children to advance for the new input symbol; if
more than one can advance, then we will have a disjunction for each
possible evolving child, with the other child graphs in each case
unchanged.  We apply the usual shift and reduce operations at leaf
nodes.  It is when a reduce operation exhausts a stack via Step~1(c)
of the algorithm that we update the controller binding,
possibly branching to multiple bindings of the controller as in the
example.  We can optimize the traversal and preserve sharing of
subgraphs by caching the map from old to new upper graph structures,
traversing a shared subgraph once only.

\section{Conclusion}

We have presented an extension to GLR parsing for languages generated
by context-free grammars enriched with the shuffle operator, and
discussed its correctness and its efficient implementation.  We
conclude with two avenues of future work.

Along with the GLR algorithm Tomita gave an approach for efficiently
representing a \emph{parse forest}, a collection of parse trees, of
all possible derivations, and we have left the adaptation of this
technique to \parseralg\ parsing to future work.  An efficient
representation of these parse forests is quite relevant to the
artificial intelligence applications of this work.  Capturing the
parse forest represents the difference between goal recognition, where
only the top-level intentions of an actor are determined, and plan
recognition, where in addition to the goal a detailed plan is
constructed.  The cost of retaining plans is not inconsiderable, and
efficient recognition of full plans remains an area of active
study~\cite{mirsky:ijcai2016}.

A formal assessment of the worst- and average-case complexity of
\parseralg\ parsing also remains to be done.  However, we have
implemented a prototype of a plan recognizer based on
\parseralg\ parsing, and our preliminary testing suggests that it
improves considerably over past approaches, with near-linear
parformance for randomly generated libraries which reflect common use
cases.

\bibliography{refs}

\begin{thebibliography}{10}

\bibitem{gallo-et-al:hypergraphs:1993}
Giorgio Gallo, Giustino Longo, Stefano Pallottino, and Sang Ngyyen.
\newblock Directed hypergraphs and applications.
\newblock {\em Discrete Applied Mathematics}, 42:177--201, 1993.

\bibitem{geib:ijcai-2009}
Christopher~W. Geib.
\newblock Delaying commitment in plan recognition using combinatory categorial
  grammars.
\newblock In {\em Proc.\ 21st Int.\ Joint Conf.\ on Artificial Intelligence},
  2009.

\bibitem{geib-goldman:ai-2009}
Christopher~W. Geib and Robert~P. Goldman.
\newblock A probabilistic plan recognition algorithm based on plan tree
  grammars.
\newblock {\em Artificial Intelligence}, 117(11):1101--1132, July 2009.

\bibitem{geib-maraist-goldman:icaps08}
Christopher~W. Geib, John Maraist, and Robert~P. Goldman.
\newblock A new probabilistic plan recognition algorithm based on string
  rewriting.
\newblock In {\em Proc.\ 18th Int.\ Conf.\ on Automated Planning and
  Scheduling}, pages 91--98, September 2008.

\bibitem{goldman-geib-miller:UAI99}
Robert~P. Goldman, Christopher~W. Geib, and Christopher~A. Miller.
\newblock A new model of plan recognition.
\newblock In {\em Proc.\ 15th Conf.\ on Uncertainty in Artificial
  Intelligence}, pages 245--254, July 1999.

\bibitem{grune-jacobs-parsing:2008}
Dick Grune and Ceriel~J.H. Jacobs.
\newblock {\em Parsing Techniques: A Practical Guide}.
\newblock Springer, 2008.

\bibitem{mirsky:ijcai2016}
Reuth Mirsky and Ya'akov Gal.
\newblock {SLIM}: Semi-lazy inference mechanism for plan recognition.
\newblock In {\em Proc.\ 25th Int.\ Joint Conf.\ on Artificial Intelligence},
  July 2016.

\bibitem{pynadath-wellman:uai-2000}
David~V. Pynadath and Michael~P. Wellman.
\newblock Probabilistic state-dependent grammars for plan recognition.
\newblock In {\em Proc. 16th Conf.\ on Uncertainty in Artificial Intelligence},
  pages 507--514. Morgan Kaufmann Publishers Inc., 2000.

\bibitem{restivo:lata2015}
Antonio Restivo.
\newblock The shuffle product: New research directions.
\newblock In {\em Proc.\ 9th Int.\ Conf.\ on Language and Automata Theory and
  Applications}, pages 70--81, March 2015.

\bibitem{sulzmann-thiemann:lata2015}
Martin Sulzmann and Peter Thiemann.
\newblock Derivatives for regular shuffle expressions.
\newblock In {\em Proc.\ 9th Int.\ Conf.\ on Language and Automata Theory and
  Applications}, March 2015.

\bibitem{sulzmann-thiemann:jcss-lata2015}
Martin Sulzmann and Peter Thiemann.
\newblock Derivatives and partial derivatives for regular shuffle expressions.
\newblock {\em Journal of Computer and System Sciences}, submitted.

\bibitem{tomita-ijcai-1985}
Masaru Tomita.
\newblock An efficient context-free parsing algorithm for natural languages.
\newblock In {\em Proc.\ 9th Int.\ Joint Conf.\ on Artificial Intelligence},
  pages 756--764, August 1985.

\bibitem{tomita-genlr-book-1991}
Masaru Tomita, editor.
\newblock {\em Generalized {LR} Parsing}.
\newblock Kluwer Academic Publishers, 1991.

\bibitem{vilain:aaai90}
Marc Vilain.
\newblock Getting serious about parsing plans: A grammatical analysis of plan
  recognition.
\newblock In {\em Proc.\ 8th Nat.\ Conf.\ on Artificial Intelligence}, pages
  190--197, 1990.

\bibitem{vilain:aaai91}
Marc Vilain.
\newblock Deduction as parsing: Tractable classification in the {KL-ONE}
  framework.
\newblock In {\em 9th Nat.\ Conf.\ on Artificial Intelligence}, pages 464--470,
  July 1991.

\end{thebibliography}
\end{document}